\documentclass[12pt]{article}
\usepackage{epsf}
\usepackage{graphicx}

\title{{\bf QCD sum rules analysis of the  rare radiative $B_{c}\rightarrow D_{s}^{\ast}\gamma $  decay }}
\author{\vspace{1cm}\\
K. Azizi$^1$ \thanks {e-mail: kazizi@newton.physics.metu.edu.tr} ,
V. Bashiry$^2$ \thanks {e-mail: bashiry@ipm.ir}\\\small $^1$
Physics Department, Middle East Technical University, 06531 \\\small
Ankara, Turkey \\\small $^2$ Institute for Studies in Theoretical
Physics and Mathematics
\\\small(IPM),
 P.O. Box 19395-5531, Tehran, Iran }
 \date{}
\begin{document}
\setlength{\baselineskip}{24pt} \maketitle
\setlength{\baselineskip}{7mm}
\begin{abstract}
 In this work, the  radiative $B_{c}\rightarrow D_{s}^{\ast}\gamma $  decay is investigated in the framework of QCD sum
rules.
  The transition form factors responsible for the decay are calculated.
  The total branching ratio for this decay is estimated to be in the order of $ 10^{-5}$, so
  this decay
  can be measurable  at LHC in the near future.
\end{abstract}
\thispagestyle{empty}
\newpage
\setcounter{page}{1}
\section{Introduction}
 The heavy pseudoscalar meson $B_{c}$
contains two heavy quarks of different flavor. This meson has been
discovered in 1998 via the decay mode $B_{c}\rightarrow J/\psi
l^{\pm}\nu$ in 1.8 TeV $p\overline{p}$ collisions, using the CDF
detector at the Fermi Lab  \cite{CDF}. The   $B_c$ meson
constitutes a very rich laboratory for studying various decay
channels.   There are three classes of decays of  $B_{c}$ meson,
namely b-quark decay (when c is spectator), c-quark decay (when b
is spectator) and the weak annihilation channels. Because of two
heavy quark contents, the $B_{c}$ decay channels are expected to
be very rich in comparison with other B mesons, so investigation
of this meson is essential from both theoretical and experimental
point of view. The $B_c$ meson decays provide windows for reliable
determination of the CKM matrix element $V_{cb}$ and can shed
light on new physics beyond the standard model.

At LHC with the luminosity values of ${\cal L}=10^{34}cm^{-2}s^{-1}$
and $\sqrt{s}=14\rm TeV$, the number of $B_c^{\pm}$ events is
expected to be about $10^{8}\sim10^{10}$ per year, so there are high
probability to study not only some rare $B_c$ decays, but also CP
violation, T violation and polarization asymmetries. Some possible
channels are $B_{c}\rightarrow l \overline{\nu}\gamma$,
$B_{c}\rightarrow \rho^{+}\gamma$, $B_{c}\rightarrow
K^{\ast+}\gamma$ and $B_{c}\rightarrow B_{u}^{\ast}l^{+}l^{-}$,
$B_{c}\rightarrow B_{u}^{\ast}\gamma $, which have been studied in
the frame of light-cone QCD and three point QCD sum rules
\cite{Aliev1, Aliev2, Aliev3, Alievsp}. A large set of exclusive
nonleptonic and semileptonic decays of the $B_{c}$ meson, which have
been studied within a relativistic constituent quark model can be
found in \cite{Ivanov}. Another possible decay channel of $B_{c}$
 is $B^-_c \to \eta' \ell^- \bar{\nu}~$ decay, which is studied
both for decay rate and lepton polarization asymmetry
\cite{bashiry}. We analyzed the radiative $B_{c}\rightarrow
D_{s}^{\ast}\gamma $  decay by using  QCD sum rules method. Using
our calculations at the end, we also analyze the $B_{c}\rightarrow
D^{\ast}\gamma $ decay  by making the necessary changes.

The main quantities in analyzing of  $B_{c}\rightarrow
D_{s}^{\ast}\gamma $ decay are the form factors. For the calculation
of form factors, relevant to this transition, we need some
nonperturbative approaches. Among the nonperturbative approaches,
QCD sum rules method has received special attention, because this
approach is based on QCD lagrangian. This method has been
successfully applied to a wide variety of problems (for a review see
\cite{Aliev1, Aliev2,Aliev3, Colangelo, Ball1, Ball2, Ovchinnikov1,
Ovchinnikov2, azizi}).

 The $B_{c}\rightarrow D_{s}^{\ast}\gamma $ decay  occurs via flavor
  changing neutral current (FCNC) transition ($b\rightarrow s\gamma$)
  and weak annihilation channels. The b quark decay (electromagnetic
penguin)  for $B_{c}\rightarrow B_{u}^{\ast}\gamma $ has been
calculated in \cite{Alievsp} (for more details about the
electromagnetic penguin diagram see also \cite{ Kiselevsp}). We
repeated the similar calculations for our problem and found that the
corresponding branching ratio contribution was less important than
that of the weak annihilation channel. Note that, the
$B_{c}\rightarrow D_{s}^{\ast}\gamma $ decay has been investigated
in the perturbative QCD (PQCD) approach \cite{Dongsheng},
relativistic independent quark model (RIQM) \cite{Barik} and in the
framework of pertubative QCD in SM (PQCD), multiscale walking
technicolor (MWTCM) and topcolor assisted MWTCM (TAMWTCM) models
\cite{Gongru}. They found  also that the contribution of the weak
annihilation was important than that of the
 electromagnetic penguin in PQCD, RIQM and TAMWTCM. In addition, $B_{c}\rightarrow D^{\ast}\gamma $
 decay has also been
investigated in the relativistic independent quark model (RIQM)
\cite{Barik}.

The paper is organized as follows: In section (2) we construct the
transition amplitude for the weak annihilation channel in terms of
four relevant form factors, where a photon can be radiated from
$B_{c}$ or $D_{s}^{\ast}$. Two of the relevant form factors for this
decay are calculated in \cite{Aliev1} in the framework of the
light--cone QCD sum rules. In section (3), we calculate the
remaining two form factors, when a photon is radiated from
$D_{s}^{\ast}$ meson also in light--cone QCD. In section (4), we
calculate the transition form factors for electromagnetic penguin in
the framework of three point QCD sum rules method. Finally, in
section (5) numerical analysis, discussion and comparison of our
results to those of the other approaches are forwarded and
conclusion is presented.
\section{Transition amplitude of the weak annihilation for the radiative $B_{c}\rightarrow D_{s}^{\ast}\gamma $ decay}

In this section, we concentrated on the main points for obtaining
the matrix elements of the radiative decay $B_{c}\rightarrow
D_{s}^{\ast}\gamma $ along the lines similar to \cite{Wyler}. The
weak annihilation mechanism for this decay is shown in (Fig.1).
\begin{figure}
\vspace*{-1cm}
\begin{center}
\includegraphics[width=10cm]{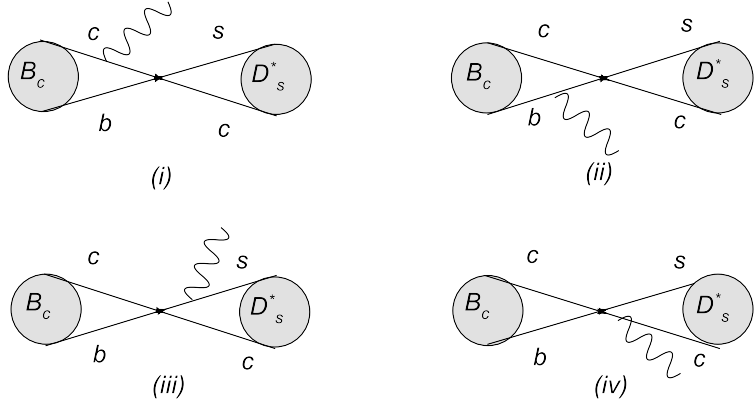}
\end{center}
\caption{The weak annihilation mechanism for $B_{c}\rightarrow
D_{s}^{\ast}\gamma $  } \label{fig1}
\end{figure}
The transition amplitude for this decay can be written as:
\begin{equation}\label{amp1}
M(B_{c}\rightarrow D_{s}^{\ast}\gamma
)=\frac{G_{F}}{\sqrt{2}}V_{cb}V_{cs}^{\ast}<D_{s}^{\ast}(p)\gamma(q)\mid(\overline{s}\Gamma_{\nu}c)(\overline{c}\Gamma^{\nu}b)\mid
B_{c}(p+q)>
\end{equation}
where $\Gamma_{\nu}=\gamma_{\nu}(1-\gamma_{5})$ and p, q and p+q are
the momentum of $D_{s}^{\ast}$, photon and $B_{c}$, respectively.
Using factorization hypothesis, The matrix element in Eq.
(\ref{amp1}) can be written in the following form:
\begin{eqnarray} \label{matrix el.}
<D_{s}^{\ast}(p)\gamma(q)\mid(\overline{s}\Gamma_{\nu}c)(\overline{c}\Gamma^{\nu}b)\mid
B_{c}(p+q)>&=&-e\varepsilon^{\mu}\varepsilon^{(D_{s}^{\ast})_{\nu}}
f_{D_{s}^{\ast}}m_{D_{s}^{\ast}}T_{\mu\nu}^{(B_{c})} \nonumber
\\&&-ie\varepsilon^{\mu}(p+q)^{\nu}
f_{B_{c}}T_{\mu\nu}^{(D_{s}^{\ast})}
\end{eqnarray}
where, the covariant decomposition of hadronic matrix elements
$T_{\mu\nu}^{(B_{c})}$ and $ T_{\mu\nu}^{(D_{s}^{\ast})}$ are
responsible for the emission of photon from initial and final
states, $f_{D_{s}^{\ast}}, ~f_{B_{c}}$ are the leptonic decay
constants of $D_{s}^{\ast}$ and $B_{c}$ mesons, respectively, and
$\varepsilon^{\mu}$ and $\varepsilon^{(D_{s}^{\ast})_{\nu}}$ are
the polarization vectors of a photon and $D_{s}^{\ast}$ meson. The
covariant decomposition of hadronic matrix elements are defined by
the following two-point correlation functions:
\begin{equation}\label{corr1}
T_{\mu\nu}^{(B_{c})}(p,q)=i\int d^{4}xe^{iqx}<0\mid
T\{j_{\mu}^{em}\overline{c}\Gamma_{\nu}b(0)\}\mid B_{c}(p+q)>
\end{equation}
and
\begin{equation}\label{corr2}
T_{\mu\nu}^{(D_{s}^{\ast})}(p,q)=i\int
d^{4}xe^{iqx}<D_{s}^{\ast}(p)\mid
T\{j_{\mu}^{em}\overline{s}\Gamma_{\nu}c(0)\}\mid 0>
\end{equation}
where $j_{\mu}^{em}$ stands for electromagnetic current. Our aim
is to construct the $T_{\mu\nu}^{(B_{c})}$ and $
~T_{\mu\nu}^{(D_{s}^{\ast})}$ in terms of form factors and other
physical quantities.  Let first focus on $T_{\mu\nu}^{(B_{c})}$.
This quantity can be written in terms of two independent 4-momenta
p and q in general as follows:
\begin{equation}\label{general1}
T_{\mu\nu}^{(B_{c})}(p,q)=g_{\mu\nu}a+p_{\mu}q_{\nu}b+q_{\mu}p_{\nu}c+p_{\mu}p_{\nu}d+q_{\mu}q_{\nu}e+
\varepsilon_{\nu\mu\lambda\sigma}p^{\lambda}q^{\sigma}F_{V}^{(B_{c})}
\end{equation}
where a, b, c, d, e and $F_{V}^{(B_{c})}$ are invariant
amplitudes. Applying the  Ward identity for electromagnetic
current to Eq. (\ref{general1}) and using the fact that for a real
photon $q^{2}=0$, we rewrite Eq.(\ref{general1}) in the following
 form:
\begin{eqnarray} \label{general2}
T_{\mu\nu}^{(B_{c})}(p,q)&=&(g_{\mu\nu}(p.q)-p_{\mu}q_{\nu})iF_{A}^{(B_{c})}+g_{\mu\nu}(p.q)\alpha+p_{\mu}q_{\nu}\beta+q_{\mu}q_{\nu}c\nonumber
\\&&+
i\frac{p_{\mu}p_{\nu}}{p.q}f_{B_{c}}+q_{\mu}q_{\nu}e+\varepsilon_{\nu\mu\lambda\sigma}p^{\lambda}q^{\sigma}F_{V}^{(B_{c})}
\end{eqnarray}
where $F_{A}^{(B_{c})}$, $\alpha$ and $\beta$ are the new
invariant amplitudes. To obtain the relation between $\alpha$ and
$\beta$, we compare Eqs. (\ref{general1}) and (\ref{general2}),
which leads to
\begin{equation}\label{abalpbet}
g_{\mu\nu}a+p_{\mu}q_{\nu}b=(g_{\mu\nu}(p.q)-p_{\mu}q_{\nu})iF_{A}^{(B_{c})}+g_{\mu\nu}(p.q)\alpha+p_{\mu}q_{\nu}\beta
\end{equation}
by multiplying both sides of Eq. (\ref{abalpbet}) with $q^{\mu}$ and
using $a+(p.q)b=if_{B_{c}}$ we get the following relation between
$\alpha$ and $\beta$, which is called the Ward identity.
\begin{equation}\label{alphabet}
\alpha+\beta=\frac{if_{B_{c}}}{(p.q)}
\end{equation}
From Eq. (\ref{alphabet}), it is clear that $\alpha$ and $\beta$
can take different choices.  Within the scop of the  present work,
in parallel with\cite{Wyler}, we set $\beta=0$ and
$\alpha=\frac{if_{B_{c}}}{(p.q)}$. Substituting the values of
$\alpha$ and $\beta$, we obtain
\begin{eqnarray} \label{general3}
T_{\mu\nu}^{(B_{c})}(p,q)&=&(g_{\mu\nu}(p.q)-p_{\mu}q_{\nu})iF_{A}^{(B_{c})}
+if_{B_{c}}g_{\mu\nu}+q_{\mu}q_{\nu}c\nonumber
\\&&+
i\frac{p_{\mu}p_{\nu}}{p.q}f_{B_{c}}+q_{\mu}q_{\nu}e+
\varepsilon_{\nu\mu\lambda\sigma}p^{\lambda}q^{\sigma}F_{V}^{(B_{c})}
\end{eqnarray}
Using $\varepsilon.q=0$,  $\varepsilon^{(D_{s}^{\ast})}.p=0$ and Eq.
(\ref{general3}), we obtain the following expression for the first
term in Eq. (\ref{matrix el.}) in terms of two form factors
($F_{A}^{(B_{c})},~F_{V}^{(B_{c})}$) :
\begin{eqnarray} \label{first term}
e\varepsilon^{\mu}\varepsilon^{(D_{s}^{\ast})_{\nu}}
f_{D_{s}^{\ast}}m_{D_{s}^{\ast}}T_{\mu\nu}^{(B_{c})}&=&ef_{D_{s}^{\ast}}m_{D_{s}^{\ast}}\{[(\varepsilon.\varepsilon^{(D_{s}^{\ast})})(p.q)
-(\varepsilon.p)(\varepsilon^{(D_{s}^{\ast})}.q)]iF_{A}^{(B_{c})}\nonumber
\\&&+if_{B_{c}}(\varepsilon.\varepsilon^{(D_{s}^{\ast})})
+\varepsilon_{\nu\mu\lambda\sigma}\varepsilon^{(D_{s}^{\ast})_{\nu}}\varepsilon^{\mu}p^{\lambda}q^{\sigma}F_{V}^{(B_{c})}\}
\end{eqnarray}
Omitting the details for the calculation
$T_{\mu\nu}^{(D_{s}^{\ast})}$, we get the following result for the
second term in the Eq. (\ref{matrix el.}):
\begin{eqnarray} \label{second term final}
ie\varepsilon^{\mu}(p+q)^{\nu}f_{B_{c}}T_{\mu\nu}^{(D_{s}^{\ast})}&=&ief_{B_{c}}\{[(\varepsilon.\varepsilon^{(D_{s}^{\ast})})(p.q)
-(\varepsilon.p)(\varepsilon^{(D_{s}^{\ast})}.q)]iF_{A}^{(D_{s}^{\ast})}\nonumber
\\&+&f_{D_{s}^{\ast}}m_{D_{s}^{\ast}}(\varepsilon.\varepsilon^{(D_{s}^{\ast})})
+\varepsilon_{\nu\mu\lambda\sigma}\varepsilon^{(D_{s}^{\ast})_{\nu}}\varepsilon^{\mu}p^{\lambda}q^{\sigma}F_{V}^{(D_{s}^{\ast})}\}~~
\end{eqnarray}
where, $F_{A}^{(D_{s}^{\ast})}$ and $F_{V}^{(D_{s}^{\ast})}$ are
two form factors of $D_s^\ast$. Now, we can write the transition
amplitude for the radiative $B_{c}\rightarrow D_{s}^{\ast}\gamma $
decay in terms of four form factors $F_{A}^{(B_{c})}$,
$F_{V}^{(B_{c})}$, $F_{A}^{(D_{s}^{\ast})}$ and
$F_{V}^{(D_{s}^{\ast})}$, as follows:
\begin{eqnarray}\label{amp2}
M(B_{c}\rightarrow D_{s}^{\ast}\gamma
)&=&e\frac{G_{F}}{\sqrt{2}}V_{cb}V_{cs}^{\ast}[-f_{D_{s}^{\ast}}m_{D_{s}^{\ast}}\{[(\varepsilon.\varepsilon^{(D_{s}^{\ast})})(p.q)
-(\varepsilon.p)(\varepsilon^{(D_{s}^{\ast})}.q)]iF_{A}^{(B_{c})}\nonumber
\\&&+if_{B_{c}}(\varepsilon.\varepsilon^{(D_{s}^{\ast})})
+\varepsilon_{\nu\mu\lambda\sigma}\varepsilon^{(D_{s}^{\ast})_{\nu}}\varepsilon^{\mu}p^{\lambda}q^{\sigma}F_{V}^{(B_{c})}\}\nonumber
\\&&-if_{B_{c}}\{[(\varepsilon.\varepsilon^{(D_{s}^{\ast})})(p.q)
-(\varepsilon.p)(\varepsilon^{(D_{s}^{\ast})}.q)]iF_{A}^{(D_{s}^{\ast})}\nonumber
\\&&+f_{D_{s}^{\ast}}m_{D_{s}^{\ast}}(\varepsilon.\varepsilon^{(D_{s}^{\ast})})
+\varepsilon_{\nu\mu\lambda\sigma}\varepsilon^{(D_{s}^{\ast})_{\nu}}\varepsilon^{\mu}p^{\lambda}q^{\sigma}F_{V}^{(D_{s}^{\ast})}\}]
\end{eqnarray}

The form factors $F_{A}^{(B_{c})}$ and $F_{V}^{(B_{c})}$
corresponding to the emission of the photon from b and c quark (see
Fig. 1 i, ii), are calculated in  \cite{Aliev1}. Therefore, we will
concentrate on the calculation of the form factors
$F_{A}^{(D_{s}^{\ast})}$ and $F_{V}^{(D_{s}^{\ast})}$ (see Fig.1
iii, iv).
%
\section{Light cone QCD sum rules for the  form factors $F_{A}^{(D_{s}^{\ast})}$ and $F_{V}^{(D_{s}^{\ast})}$  }

Based on the general idea on QCD sum rules method, we will calculate
the transition form factors by equating the representation of a
suitable correlator in hadronic and quark--gluon languages. For this
aim, we consider the following correlation function:
\begin{equation}\label{corr.4}
\Pi_{\mu\nu}(p,q)=i\int d^{4}xe^{iQx}<\gamma(q)\mid
T\{\overline{c}(x)\gamma_{\mu}(1-\gamma_{5})s(x)\overline{s}(0)\gamma_{\nu}c(0)\}\mid
 0>
\end{equation}
Here, q and p are the momentum values of photon and
$D_{s}^{\ast}$, respectively and $Q=p+q$ is the transferred
momentum. Now, we insert the hadronic state $D_{s}^{\ast}(p)$ to
Eq. (\ref{corr.4}). This can be re-written as:
\begin{equation}\label{corr.5}
\Pi_{\mu\nu}(p,q)=\frac{<\gamma(q)\mid
\overline{c}\gamma_{\mu}(1-\gamma_{5})s\mid
D_{s}^{\ast}(p)><D_{s}^{\ast}(p)\mid\overline{s}\gamma_{\nu}c\}\mid
 0>}{m_{D_{s}^{\ast}}^{2}-p^{2}}.
\end{equation}
The second term in Eq. (\ref{corr.5}), by definition is:
\begin{equation}\label{dlepcos.}
<D_{s}^{\ast}(p)\mid\overline{s}\gamma_{\nu}c\mid
 0>=f_{D_{s}^{\ast}}m_{D_{s}^{\ast}}\varepsilon_{\nu}^{(D_{s}^{\ast})}
\end{equation}
From Fig. 1 (iii, iv) and due to the fact that parity, Lorentz
 and gauge invariance are musts.  We can write the matrix element for the
emission of the photon from $D_{s}^{\ast}$ meson as:
\begin{eqnarray}\label{definition}
<\gamma(q)\mid\overline{c}\gamma_{\mu}(1-\gamma_{5})s\mid
 D_{s}^{\ast}(p)>&=&e\{i\varepsilon_{\mu\alpha\beta\sigma}\varepsilon^{\alpha}\varepsilon^{(D_{s}^{\ast})_{\beta}}q^{\sigma}
 \frac{F_{V}^{(D_{s}^{\ast})}(Q^{2})}{m_{D_{s}^{\ast}}^{2}}\nonumber
\\&+&[\varepsilon_{\mu}(\varepsilon^{(D_{s}^{\ast})}.q)-(\varepsilon.
 \varepsilon^{(D_{s}^{\ast})})q_{\mu}]\frac{F_{A}^{(D_{s}^{\ast})}(Q^{2})}{m_{D_{s}^{\ast}}^{2}}\}~~~~~
\end{eqnarray}
Substituting Eqs. (\ref{dlepcos.}) and (\ref{definition}) to
(\ref{corr.5}), we have
\begin{eqnarray}\label{corr.6}
\Pi_{\mu\nu}(p,q)&=&\frac{ef_{D_{s}^{\ast}}m_{D_{s}^{\ast}}}{m_{D_{s}^{\ast}}^{2}-p^{2}}\{i\varepsilon_{\mu\alpha\beta\sigma}\varepsilon^{\alpha}
\varepsilon_{\nu}^{(D_{s}^{\ast})}\varepsilon^{(D_{s}^{\ast})_{\beta}}q^{\sigma}
 \frac{F_{V}^{(D_{s}^{\ast})}(Q^{2})}{m_{D_{s}^{\ast}}^{2}}\nonumber
\\&+&[\varepsilon_{\mu}\varepsilon_{\nu}^{(D_{s}^{\ast})}(\varepsilon^{(D_{s}^{\ast})}.q)-(\varepsilon.
 \varepsilon^{(D_{s}^{\ast})})\varepsilon_{\nu}^{(D_{s}^{\ast})}q_{\mu}]\frac{F_{A}^{(D_{s}^{\ast})}(Q^{2})}{m_{D_{s}^{\ast}}^{2}}\}
\end{eqnarray}
Summation over polarization of $D_{s}^{\ast}$ meson is performed
by using:
\begin{equation}\label{mult}
\varepsilon_{\nu}^{(D_{s}^{\ast})}\varepsilon_{\beta}^{(D_{s}^{\ast})}=-g_{\nu\beta}+\frac{p_{\nu}p_{\beta}}{m_{D_{s}^{\ast}}^{2}}
\end{equation}
After performing the standard calculations for the
phenomenological part, we get:
\begin{eqnarray}\label{corr.7}
\Pi_{\mu\nu}(p,q)=\frac{ef_{D_{s}^{\ast}}m_{D_{s}^{\ast}}}{m_{D_{s}^{\ast}}^{2}-p^{2}}\{i\varepsilon_{\mu\nu\alpha\sigma}\varepsilon^{\alpha}
q^{\sigma}
 \frac{F_{V}^{(D_{s}^{\ast})}(Q^{2})}{m_{D_{s}^{\ast}}^{2}}+[q_{\mu}\varepsilon_{\nu}-\varepsilon_{\mu}q_{\nu}]
 \frac{F_{A}^{(D_{s}^{\ast})}(Q^{2})}{m_{D_{s}^{\ast}}^{2}}\}
\end{eqnarray}

The theoretical part (QCD side) of the correlator is calculated by
means of OPE up to operators having dimension $d=5$ in deep
Euclidean space, where both $p^{2}$ and $Q^{2}$ are large and
negative. It is determined by the bare--loop (fig. 2(a, b)) and the
power corrections from the operators with $d=3$,
$<\overline{\psi}\psi>$, $d=4$, $m_{s}<\overline{\psi}\psi>$, $d=5$,
$m_{0}^{2}<\overline{\psi}\psi>$ (fig. 2(c, d, e)) and the photon
interaction with a soft quark line (fig. 2f).
 In calculating the
bare-loop and nonperturbative correction contributions, we first
write the Lorentz decomposition of the correlator as:
\begin{eqnarray}\label{decompositionm}
\Pi_{\mu\nu}(p,q)=i\varepsilon_{\mu\nu\alpha\sigma}\varepsilon^{\alpha}
q^{\sigma}
 \Pi_{1}+[q_{\mu}\varepsilon_{\nu}-\varepsilon_{\mu}q_{\nu}]
 \Pi_{2}
\end{eqnarray}
and the dispersion representation (Cutkosky method) for the
coefficients of corresponding Lorentz structures appearing in the
$\Pi_{\mu\nu}(p,q)$ as follows:
\begin{eqnarray}\label{T12}
\Pi_{1,2}(p,q)=\int
ds\frac{\rho_{1,2}(s,p^{2})}{s-Q^{2}}+\mbox{subtraction terms}
\end{eqnarray}
where $\rho_{1, 2}(s, p^{2})$ are spectral density corresponding
to two structures in $\Pi_{\mu\nu}(p,q)$ and subtraction terms
stand for corrections. To calculate $\rho_{1,2}(s,p^{2})$, we
consider Feynmen diagrams in Fig. 2(a, b).
\begin{figure}
\vspace*{-1cm}
\begin{center}
\includegraphics[width=10cm]{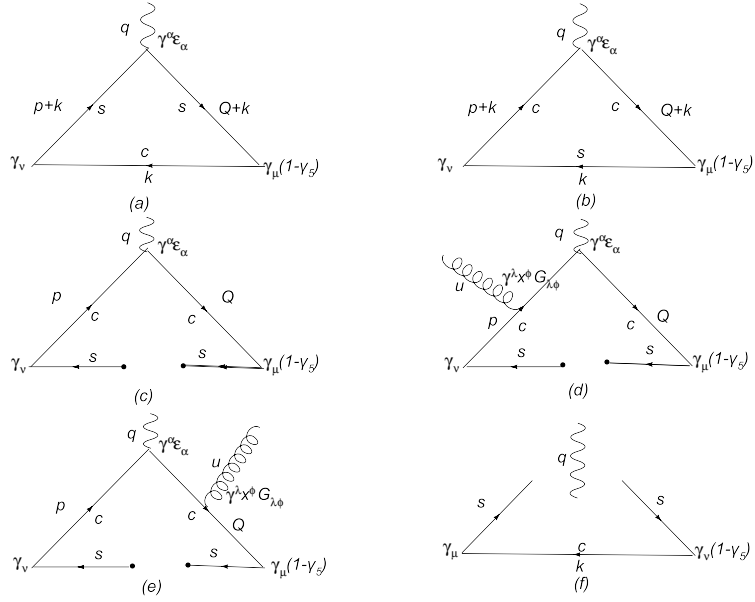}
\end{center}
\caption{Feynmen diagrams for bare-loop (a, b), power corrections
from the operators with 3, 4 and 5 dimensions (c, d, e) and
propagation of the soft quark in electromagnetic field (f) }
\label{fig,2}
\end{figure}
For instance, for the contribution of diagram (a) we get
\begin{eqnarray}\label{diagram a}
T_{\mu\nu}&=&eN_{c}Q_{s}\int\frac{d^{4}k}{(2\pi)^{4}}\nonumber
\\&&\{Tr[\frac{i(\not\!k+m_{c})}{k^{2}-m_{c}^{2}}\gamma_{\mu}(1-\gamma_{5})\frac{i(\not\!Q+\not\!k+m_{s})}{(Q+k)^{2}
-m_{s}^{2}}\not\!\varepsilon\frac{i(\not\!p+\not\!k+m_{s})}{(p+k)^{2}
-m_{s}^{2}}\gamma_{\nu}]\}\nonumber
\\
\end{eqnarray}
 With the help of the above equations, we obtain the following expressions
corresponding to the coefficients of the structures
$i\varepsilon_{\mu\nu\alpha\sigma}\varepsilon^{\alpha} q^{\sigma}$
and $[q_{\mu}\varepsilon_{\nu}-\varepsilon_{\mu}q_{\nu}]$:
\begin{eqnarray}\label{(T1T2)2}
T_{1}&=&\frac{-eN_{c}Q_{s}}{4\pi^{2}}\{\int_{0}^{1}dx
x\int_{0}^{1}dy[m_{c}m_{s}-m_{s}^2x\overline{y}+p^{2}(X_{1}\overline{y}
+X_{3})-(p.q)(X_{2}\overline{y}^{2}\nonumber
\\&-&4x^{2}\overline{x}y+2x^{2}\overline{x}y^{2}-4x\overline{x}y)]\int_{0}^{\infty}d\alpha e^{-\alpha\Delta}\},\nonumber
\\
T_{2}&=&\frac{eN_{c}Q_{s}}{16\pi^{2}}\{\int_{0}^{1}dx
x\int_{0}^{1}dy[m_{c}m_{s}-m_{s}^2x\overline{y}+p^{2}(X_{1}\overline{y}
-X_{3})-(p.q)X_{3}(2\overline{y}\nonumber
\\&+&4\overline{x}y)]\int_{0}^{\infty}d\alpha
e^{-\alpha\Delta}\}
\end{eqnarray}
where $\overline{x}(\overline{y})=1-x(y)$,
$\Delta=-p^{2}x\overline{x}y-Q^{2}x\overline{x}~\overline{y}+
m_{c}^{2}\overline{x}+m_{s}^{2}x$, $X_{1}=x-2x^{2}+x^{3}$,
$X_{2}=x^{2}-2x^{3}$, $X_{3}=-x+x^{2}$

In this calculation,  we have also used the exponential
representation for the denominator as:
\begin{eqnarray}\label{denominator}
\frac{1}{\Delta^{n}}=\frac{1}{(n-1)!}\int_{0}^{\infty}d\alpha
\alpha^{n-1} e^{-\alpha\Delta}
\end{eqnarray}
Next, we apply the double Borel operator
$\hat{B}(M_{1}^{2})\hat{B}(M_{2}^{2})$ on $T_{(1,2)}$ and we get:
\begin{eqnarray}\label{(T1T2)3}
\tilde{T_{1}}&=&\frac{-eN_{c}Q_{s}}{4\pi^{2}}\frac{\sigma_{1}\sigma_{2}}{\sigma_{1}+\sigma_{2}}\int_{0}^{1}dx
\frac{1}{\overline{x}}e^{\frac{1}{x\overline{x}}(m_{c}^{2}\overline{x}+m_{s}^{2}x)(\sigma_{1}+\sigma_{2})}[m_{c}m_{s}-m_{s}^2x
\frac{\sigma_{2}}{\sigma_{1}+\sigma_{2}}\nonumber
\\&+&p^{2}((x^{3}-2x^{2})\frac{\sigma_{2}}{\sigma_{1}+\sigma_{2}}+x^{2}-x\frac{\sigma_{1}}{\sigma_{1}
+\sigma_{2}})-(p.q)(x^{2}-2x^{3}+(3x^{2}\nonumber
\\&-&4x^{3})\frac{\sigma_{1}^{2}}{(\sigma_{1}+\sigma_{2})^{2}}+(8x^{3}-2x^{2}-4x)\frac{\sigma_{1}}{\sigma_{1}
+\sigma_{2}})],\nonumber
\\
\tilde{T_{2}}&=&\frac{eN_{c}Q_{s}}{16\pi^{2}}\frac{\sigma_{1}\sigma_{2}}{\sigma_{1}+\sigma_{2}}\int_{0}^{1}dx
\frac{1}{\overline{x}}e^{\frac{1}{x\overline{x}}(m_{c}^{2}\overline{x}+m_{s}^{2}x)(\sigma_{1}+\sigma_{2})}[m_{c}m_{s}-m_{s}^2x
\frac{\sigma_{2}}{\sigma_{1}+\sigma_{2}}\nonumber
\\&+&p^{2}((x+x^{3}-2x^{2})\frac{\sigma_{2}}{\sigma_{1}+\sigma_{2}}-x^{2}+x)-(p.q)(-2x+2x^{2})\frac{\sigma_{2}}{\sigma_{1}+\sigma_{2}}
\nonumber
\\&-&2x^{3}\frac{\sigma_{2}^{2}}{(\sigma_{1}+\sigma_{2})^{2}}+(8x^{2}-4x^{3}-4x)\frac{\sigma_{1}}{\sigma_{1}
+\sigma_{2}})+(2x^{3}-2x^{2})\frac{\sigma_{1}^{2}}{(\sigma_{1}+\sigma_{2})^{2}}]\nonumber
\\
\end{eqnarray}
where $\sigma_{1}=\frac{1}{M_{1}^{2}}$ and
$\sigma_{2}=\frac{1}{M_{2}^{2}}$.

In deriving Eq. (\ref{(T1T2)3}), we  use the definition
\begin{equation}\label{Borel}
\hat{B}(M^{2})e^{-\alpha p^{2}}=\delta(1-\alpha M^{2})
\end{equation}
For the determination of the spectral density, we apply the Borel
transformations to $\tilde{T_{1}}$ and $\tilde{T_{2}}$
\cite{Nesterenko} and we obtain:
\begin{equation}\label{spectral}
\varrho_{1,2}(s,t,p^{2})=\frac{1}{st}\hat{B}(\frac{1}{s},\sigma_{1})\hat{B}(\frac{1}{t},\sigma_{2})\frac{\tilde{T_{1,2}}}{\sigma_{1}\sigma_{2}}
\end{equation}
In this step, we use the following relations:
\begin{equation}\label{sigma}
\sigma^{n}
e^{-\alpha\sigma}=(-\frac{d}{d\alpha})^{n}e^{-\alpha\sigma},
\end{equation}
\begin{equation}\label{our spec}
\rho_{1,2}(s,p^{2})=\int dt\frac{\varrho_{1,2}(s,t,p^{2})}{t-p^{2}},
\end{equation}
and
\begin{equation}\label{delta}
\hat{B}(\frac{1}{s},\sigma_{1})\hat{B}(\frac{1}{t},\sigma_{2})e^{-\alpha(\sigma_{1}
+\sigma_{2})}=\delta(1-\frac{\alpha}{s})\delta(1-\frac{\alpha}{t})
\end{equation}
 Then, we get the following expressions for
the two spectral densities, as follows:
\begin{eqnarray}\label{spec.int.form}
\rho_{1}(s,p^{2})&=&\frac{eN_{c}Q_{s}}{4\pi^2}\frac{1}{(s-p^2)^{3}}\int_{x_{0}}^{x_{1}}dx\frac{1}{\overline{x}}\{(m_{c}m_{s}-m_{s}^2x)(s-p^{2})^{2}+p^{2}
[(x^{3}-x^{2})\nonumber
\\&&(s-p^{2})^{2}-x(p^{2}+\frac{m_{c}^{2}}{x}+\frac{m_{s}^{2}}{\overline{x}})(s-p^{2})]-\frac{1}{2}(m_{s}^{2}-p^{2})[(x^{2}-2x^{3})
\nonumber
\\&&(s-p^{2})^{2}-2(3x^{2}-4x^{3})(p^{2}+\frac{m_{c}^{2}}{x}+\frac{m_{s}^{2}}{\overline{x}})^{2}+(8x^{3}-2x^{2}-4x)(p^{2}\nonumber
\\&&+\frac{m_{c}^{2}}{x}+\frac{m_{s}^{2}}{\overline{x}})
(s-p^{2})]\},\nonumber
\\
\rho_{2}(s,p^{2})&=&\frac{eN_{c}Q_{s}}{16\pi^2}\frac{1}{(s-p^2)^{3}}\int_{x_{0}}^{x_{1}}dx\frac{1}{\overline{x}}\{[m_{c}m_{s}-m_{s}^2x+p^{2}
(2x-3x^{2}+x^{3})]\nonumber
\\&&(s-p^{2})^{2}
-\frac{1}{2}(m_{s}^{2}-p^{2})[(-2x+2x^{2}-4x^{3})
(s-p^{2})^{2}+(8x^{2}-4x^{3}\nonumber
\\&&-4x)(p^{2}+\frac{m_{c}^{2}}{x}+\frac{m_{s}^{2}}{\overline{x}})(s-p^{2})-2(2x^{3}-2x^{2})(p^{2}+\frac{m_{c}^{2}}{x}+\frac{m_{s}^{2}}{\overline{x}})^{2}
]\}\nonumber
\\
\end{eqnarray}
where the integration region is determined by the following
inequality:
\begin{equation}\label{inequality}
s x \overline{x} - (m_{c}^2\overline{x}  + m_{s}^2 x)\geq0
\end{equation}
 Similar to above
calculations for diagram (a), we have repeated the entire
calculations for diagram (b). Finally, we get the following
results for the spectral densities:
\begin{eqnarray}\label{finalspecdens}
\rho_{1}(s,p^{2})&=&\frac{eN_{c}}{48\pi^2}\frac{1}{(s-p^2)^{3}}\{
Q_{s} [\lambda\{3(-3+ 5\alpha-5 \beta)p^{6}+(28 + \alpha (-103
\nonumber
\\&&+ \alpha (17 + 4 \alpha))+52 \beta-\alpha (61 + 8 \alpha) \beta + 4(10 +
\alpha)\beta^{2})p^{4}s-3 (-2 \nonumber
\\&&+ 16\alpha^{3}+\beta -
31\beta^{2}-\alpha^{2} (75 + 32 \beta) + \alpha (39 + 2 \beta (41 +
8 \beta))+7\lambda^{2}\nonumber
\\&&)p^{2}s^{2}+\alpha (1 - 86\alpha^{2}+\beta-86\beta^{2}+\alpha (67 + 172 \beta) + 18
\lambda^{2})s^{3}\}+6 \{\nonumber
\\&&p^{6}+2 m_{c} m_{s} (s-p^2)^{2}+(2 -
11 \alpha + 4 \beta)p^{4}s(-1 + 26\alpha^{2}+2\beta^{2}\nonumber
\\&&-2 \alpha (7 + 11
\beta))p^2s^2-\alpha (1 + 18\alpha^{2}+2 \beta (1 + \beta) - 4
\alpha (4 + 5 \beta))s^{3}\}\nonumber
\\&&ln\frac{1+\alpha-\beta-\lambda}{1+\alpha-\beta+\lambda}]+Q_{c} [\lambda\{3(-3+ 5\beta-5 \alpha
)p^{6}+(28 +
\beta (-103 \nonumber
\\&&+ \beta (17 + 4 \beta))+52 \alpha-\beta (61 + 8 \beta) \alpha + 4(10 +
\beta)\alpha^{2})p^{4}s-3 (-2 \nonumber
\\&&+ 16\beta^{3}+\alpha -
31\alpha^{2}-\beta^{2} (75 + 32 \alpha) + \beta (39 + 2 \alpha(41 +
8 \alpha))+7\lambda^{2}\nonumber
\\&&)p^{2}s^{2}+\beta (1 - 86\beta^{2}+\alpha-86\alpha^{2}+\beta (67 + 172 \alpha) + 18
\lambda^{2})s^{3}\}+6 \{\nonumber
\\&&p^{6}+2 m_{c} m_{s} (s-p^2)^{2}+(2 -
11 \beta + 4 \alpha)p^{4}s(-1 + 26\beta^{2}+2\alpha^{2}\nonumber
\\&&-2 \beta (7 + 11
\alpha))p^2s^2-\beta (1 + 18\beta^{2}+2 \alpha (1 + \alpha) - 4
\beta (4 + 5 \alpha))s^{3}\}\nonumber
\\&&ln\frac{1+\beta-\tilde{\alpha}-\lambda}{1+\beta-\alpha+\lambda}]\},\nonumber
\\
\rho_{2}(s,p^{2})&=&\frac{eN_{c}}{96\pi^2}\frac{1}{(s-p^2)^{3}}\{
Q_{s} [\lambda \{-(29 +2\alpha^{2}+\beta (17 + 2 \beta) - \alpha (19
+ 4 \beta))p^{6}\nonumber
\\&&+(4\alpha^{3}+\alpha (5 + \beta) (5 + 4
\beta) -\alpha^{2} (23 + 8 \beta) + 6 (8 + \beta
-\beta^{2}))p^{4}s\nonumber
\\&&+(-23 +2\alpha^{3}-5 \beta (1 + 4 \beta) + \alpha(-9 + 2 \beta (21 + \beta)) +\alpha^{2} (-30\nonumber
\\&& - 4 \beta+\frac{48}
{\lambda^{2}-(1 + \alpha - \beta)^{2}} ))p^{2}s^{2}+ \alpha (13 +
\alpha (-5 + 22  \alpha) + \beta \nonumber
\\&&- 44  \alpha \beta + 22 \beta^{2})s^{3}\}+6\{-2p^{6}+m_{c} m_{s} (s-p^2)^{2}+(4 + 5 \alpha)p^{4}s\nonumber
\\&&-2 (1 + \alpha
 + 4\alpha^{2}-2 \alpha \beta)p^{2}s^{2}+\alpha (1 + 4 \alpha (\alpha -
 \beta))s^{3}\}\nonumber
\\&&ln\frac{1-\alpha+\beta-\lambda}{1-\alpha+\beta+\lambda}]+Q_{c} [\lambda
 \{-(29 +2\beta^{2}+\alpha (17 + 2
\alpha) - \beta (19 + 4 \alpha))\nonumber
\\&&p^{6}+(4\beta^{3}+\beta (5 + \alpha) (5 + 4
\alpha) -\beta^{2} (23 + 8 \alpha) + 6 (8 + \alpha
-\alpha^{2}))p^{4}s\nonumber
\\&&+(-23 +2\beta^{3}-5 \alpha (1 + 4 \alpha) + \beta(-9 + 2 \alpha (21 + \alpha)) +\beta^{2} (-30\nonumber
\\&& - 4 \alpha+\frac{48}
{\lambda^{2}-(1 + \beta - \alpha)^{2}} ))p^{2}s^{2}+ \beta (13 +
\beta (-5 + 22  \beta) + \alpha \nonumber
\\&&- 44  \beta \alpha + 22 \alpha^{2})s^{3}\}+6\{-2p^{6}+m_{c} m_{s} (s-p^2)^{2}+(4 + 5 \beta)p^{4}s\nonumber
\\&&-2 (1 + \beta
 + 4\beta^{2}-2 \beta \alpha)p^{2}s^{2}+\beta (1 + 4 \beta (\beta -
 \alpha))s^{3}\}\nonumber
\\&&ln\frac{1-\beta+\alpha-\lambda}{1-\beta+\alpha+\lambda}]\}
 \end{eqnarray}
 where $\alpha=\frac{m_{s}^{2}}{s}$ and
 $\lambda=\sqrt{1+\alpha^{2}+\beta^{2}-2\alpha-2\beta-2\alpha\beta}$.

 The next step is to calculate contributions coming from the power
 correction
 terms. After standard but lengthy calculations for the contributions of the diagrams (c, d, e
 ), we get:
 \begin{eqnarray}\label{diagc,d,e}
\Pi_{1}(p,q)^{(c, d,
 e)}&=&\frac{m_{c}}{r'r}<\overline{s}s>-\frac{m_{s}}{2}<\overline{s}s>[\frac{m_{c}^{2}}{r'^{2}r}+\frac{1}{r'r}+\frac{m_{c}^{2}}{r'r^{2}}]+
 \frac{m_{s}^{2}}{2}<\overline{s}s>[\frac{2m_{c}^{3}}{r'^{3}r}\nonumber
\\&&+\frac{2m_{c}^{3}}{r'^{2}r^{2}}+\frac{m_{c}}{r'^{2}r}+\frac{m_{c}}{r'r^{2}}
 +\frac{2m_{c}^{3}}{r'r^{3}}]-
 \frac{m_{0}^{2}}{4}<\overline{s}s>[\frac{2m_{c}^{3}}{r'^{3}r}+\frac{2m_{c}^{3}}{r'^{2}r^{2}}\nonumber
\\&&+\frac{2m_{c}}{3r'^{2}r}+\frac{m_{c}}{r'r^{2}}
 +\frac{2m_{c}^{3}}{r'r^{3}}],\nonumber
\\
\Pi_{2}(p,q)^{(c, d,
 e)}&=&\frac{-m_{c}}{r'r}<\overline{s}s>+\frac{m_{s}}{2}<\overline{s}s>[\frac{m_{c}^{2}}{r'^{2}r}+\frac{m_{c}^{2}}{r'r^{2}}]-
 \frac{m_{s}^{2}}{2}<\overline{s}s>[\frac{2m_{c}^{3}}{r'^{3}r}\nonumber
\\&&+\frac{2m_{c}^{3}}{r'^{2}r^{2}}+\frac{m_{c}}{r'^{2}r}+\frac{m_{c}}{r'r^{2}}
 +\frac{2m_{c}^{3}}{r'r^{3}}]+
 \frac{m_{0}^{2}}{4}<\overline{s}s>[\frac{2m_{c}^{3}}{r'^{3}r}+\frac{2m_{c}^{3}}{r'^{2}r^{2}}\nonumber
\\&&+\frac{2m_{c}}{3r'^{2}r}+\frac{m_{c}}{r'r^{2}}
 +\frac{2m_{c}^{3}}{r'r^{3}}]
\end{eqnarray}
where $r^{2}=p^{2}-m_{c}^{2}$ and $r'^{2}=Q^{2}-m_{c}^{2}$.
Finally, we calculate the contribution of diagram (f). For the
calculation of this diagram corresponding to the propagation of
the soft quark in the external electromagnetic field, we use the
light-cone expansion for the non--local operators. After
contracting the c quark lines in
\begin{equation}\label{fdiag}
\Pi_{\mu\nu}(p,q)=i\int d^{4}xe^{iQx}<\gamma(q)\mid
T\{\overline{s}(0)\gamma_{\mu}c(0)\overline{c}(x)\gamma_{\nu}(1-\gamma_{5})s(x)\}\mid
 0>
\end{equation}
we obtain
\begin{equation}\label{contractingc}
\Pi_{\mu\nu}(p,q)=i\int
d^{4}x\frac{d^{4}k}{(2\pi)^{2}}\frac{e^{i(Q-k)x}}{m_{c}^{2}-k^{2}}<\gamma(q)\mid
\overline{s}\gamma_{\mu}(\not\!k+m_{c})\gamma_{\nu}(1-\gamma_{5})s\}\mid
 0>
\end{equation}
To calculate the matrix element appearing in the above equation,
we use the following identities:
 \begin{eqnarray}\label{identities}
\gamma_{\mu}\gamma_{\nu}&=&g_{\mu\nu}+i\sigma_{\mu\nu},\nonumber
\\
\gamma_{\mu}\gamma_{\nu}\gamma_{5}&=&g_{\mu\nu}\gamma_{5}-\frac{i}{2}\varepsilon_{\mu\nu\alpha\sigma}\sigma_{\alpha\sigma},\nonumber
\\
\gamma_{\mu}\gamma_{\alpha}\gamma_{\nu}&=&g_{\mu\alpha}\gamma_{\nu}+g_{\alpha\nu}\gamma_{\nu}-g_{\mu\nu}\gamma_{\alpha}
+i\varepsilon_{\mu\nu\alpha\sigma}\gamma_{\sigma}\gamma_{5}
\end{eqnarray}
and photon distribution amplitudes (DA's) for twist 2, 3 and 4
\cite{Rohrwild, Ball3}:
\begin{eqnarray}\label{DA's}
<\gamma(q)\mid \overline{s}\gamma_{\nu}s\mid
 0>&=&-\frac{Q_{s}}{2}f_{3\gamma}\int_{0}^{1}du\overline{\psi}^{(V)}(u)x^{\theta}F_{\theta\nu}(ux)\nonumber
\\
 <\gamma(q)\mid \overline{s}\gamma_{\alpha}\gamma_{5}s\mid
 0>&=&-\frac{iQ_{s}}{4}f_{3\gamma}\int_{0}^{1}du\psi^{(A)}(u)x^{\theta}\tilde{F_{\theta\alpha}}(ux)\nonumber
\\
<\gamma(q)\mid \overline{s}\sigma_{\alpha\beta}s\mid
 0>&=&Q_{s}<\overline{s}s>\int_{0}^{1}du\phi(u)F_{\alpha\beta}(ux)\nonumber
\\&&+\frac{Q_{s}<\overline{s}s>}{16}\int_{0}^{1}dux^{2}\mathcal{A}(u)
 F_{\alpha\beta}(ux)\nonumber
\\&&+\frac{Q_{s}<\overline{s}s>}{8}\int_{0}^{1}du\mathcal{B}(u)x^{\rho}(x_{\beta}F_{\alpha\rho}(ux)-x_{\alpha}F_{\beta\rho}(ux))
 \nonumber
\\
\end{eqnarray}
where $F_{\mu\nu}$ is the field strength tensor of the
electromagnetic field, which is defined by
\begin{eqnarray}\label{field strength}
F_{\mu\nu}(x)&=&i(\varepsilon_{\nu}q_{\mu}-\varepsilon_{\mu}q_{\nu})e^{iqx}\nonumber
\\
\tilde{F_{\mu\nu}}(x)&=&\frac{1}{2}\varepsilon_{\mu\nu\alpha\beta}F_{\alpha\beta}(x)
\end{eqnarray}

The asymptotic expression for the photon wave function $\phi(u)$
in terms of magnetic susceptibility of the quark condensate,
$\chi(\mu)$, at a re--normalization scale ($\mu=1 ~GeV^{2}$) is
defined by:
\begin{equation}\label{sus}
\phi(u)=\chi(\mu)u(1-u)
\end{equation}
Other functions used in Eq. (\ref{DA's}) are defined by
\cite{Rohrwild, Ball3}
\begin{eqnarray}\label{other}
\overline{\psi}^{(V)}(u)&=&-20u(1-u)(2u-1)+\frac{15}{16}(\omega_{\gamma}^{A}-3\omega_{\gamma}^{V})u(1-u)(2u-1)\nonumber
\\&&\times(7(2u-1)^{2}-3),\nonumber
\\
\psi^{(A)}(u)&=&(1-(2u-1)^{2})(5(2u-1)^{2}-1)\times\frac{5}{2}(1+\frac{9}{16}\omega_{\gamma}^{V}-\frac{3}{16}\omega_{\gamma}^{A}),\nonumber
\\
\mathcal{A}(u)&=&
40u(1-u)(3k-k^{+}+1)+8(\xi_{2}^{+}-3\xi_{2})[u(1-u)(2+13u(1-u))\nonumber
\\&&+2u^{3}(10-15u+6u^{2})lnu+2(1-u)^{3}(10-15(1-u)+6(1\nonumber
\\&&-u^{2}))ln(1-u)],\nonumber
\\
\mathcal{B}(u)&=&40\int_{0}^{u}d\alpha(4-\alpha)(1+3k^{+})[\frac{-1}{2}+\frac{3}{2}(2\alpha-1)^{2}]
\end{eqnarray}
where $k,~k^{+},~\xi_{2}$, $\xi_{2}^{+}$ and $f_{3\gamma}$ are
constants (see \cite{Rohrwild, Ball3}). Using the above relations
in Eq. (\ref{contractingc}), we obtain:
\begin{eqnarray}\label{replacement}
\Pi_{\mu\nu}(p,q)&=&-\int
d^{4}x\frac{d^{4}k}{(2\pi)^{2}}\frac{e^{i(p-k)x}}{m_{c}^{2}-k^{2}}\{\frac{-Q_{s}}{2}f_{3\gamma}k_{\mu}\int_{0}^{1}du\overline{\psi}^{(V)}
(u)x^{\sigma}F_{\sigma\nu}-\frac{Q_{s}}{2}f_{3\gamma}k_{\nu}\nonumber
\\&&\times\int_{0}^{1}du\overline{\psi}^{(V)}
(u)x^{\sigma}F_{\sigma\mu}-\frac{Q_{s}}{2}f_{3\gamma}g_{\mu\nu}k_{\alpha}\int_{0}^{1}du\overline{\psi}^{(V)}
(u)x^{\sigma}F_{\sigma\alpha}+\frac{Q_{s}}{4}f_{3\gamma}\nonumber
\\&&\times k_{\alpha}\int_{0}^{1}du\psi^{(A)}
(u)[-x^{\alpha}F_{\mu\nu}+x_{\mu}F_{\alpha\nu}+x_{\nu}F_{\mu\alpha}]+\frac{iQ_{s}}{8}f_{3\gamma}[k_{\mu}\varepsilon_{\theta\nu\eta\lambda}
\nonumber
\\&&+k_{\nu}\varepsilon_{\theta\mu\eta\lambda}-g_{\mu\nu}k_{\alpha}\varepsilon_{\theta\alpha\eta\lambda}]
\int_{0}^{1}du\psi^{(A)}
(u)x^{\theta}F_{\eta\lambda}+\frac{iQ_{s}}{2}f_{3\gamma}k_{\alpha}\varepsilon_{\mu\nu\alpha\sigma}\nonumber
\\&&\int_{0}^{1}du\overline{\psi}^{(V)}
(u)x^{\theta}F_{\theta\sigma}+im_{c}[Q_{s}<\overline{s}s>\int_{0}^{1}du\phi(u)F_{\mu\nu}(ux)\nonumber
\\&&+\frac{Q_{s}<\overline{s}s>}{16}
\int_{0}^{1}dux^{2}\mathcal{A}(u)
 F_{\mu\nu}(ux)+\frac{Q_{s}<\overline{s}s>}{8}\int_{0}^{1}du\mathcal{B}(u)x^{\sigma}\nonumber
\\&&\times(x_{\nu}F_{\mu\sigma}(ux)-x_{\mu}F_{\nu\sigma}(ux))]+\frac{m_{c}}{2}Q_{s}<\overline{s}s>\varepsilon_{\mu\nu\alpha\sigma}
[\int_{0}^{1}du\phi(u)\nonumber
\\&&\times F_{\alpha\sigma}(ux)+\frac{1}{16}\int_{0}^{1}dux^{2}\mathcal{A}(u)
 F_{\alpha\sigma}(ux)+\frac{1}{8}\int_{0}^{1}du\mathcal{B}(u)x^{\theta}
(x_{\sigma}F_{\alpha\theta}(ux)\nonumber
\\&&-x_{\alpha}F_{\sigma\theta}(ux))]\}
\end{eqnarray}
After performing integration over x and k, the following results
corresponding to the coefficients of two invariant structures
$i\varepsilon_{\mu\nu\alpha\sigma}\varepsilon^{\alpha} q^{\sigma}$
and $[q_{\mu}\varepsilon_{\nu}-\varepsilon_{\mu}q_{\nu}]$  are
obtained as follows:
\begin{eqnarray}\label{diagf}
\Pi_{1}(p,q)^{(f)}&=&\frac{m_{c}Q_{s}<\overline{s}s>}{2}[\int_{0}^{1}du\phi(u)\frac{1}{p^{2}-m_{c}^{2}}-
\frac{1}{16}\int_{0}^{1}du\mathcal{A}(u)(\frac{10}{(p^{2}-m_{c}^{2})^{2}}\nonumber
\\&&+\frac{8m_{c}^{2}}{(p^{2}-m_{c}^{2})^{3}})],\nonumber
\\
\Pi_{2}(p,q)^{(f)}&=&\frac{-Q_{s}}{4}f_{3\gamma}\int_{0}^{1}du\psi^{(A)}(u)(\frac{1}{p^{2}-m_{c}^{2}}+\frac{2m_{c}^{2}}{(p^{2}-m_{c}^{2})^{2}})
-m_{c}Q_{s}<\overline{s}s>\nonumber
\\&&[\int_{0}^{1}du\phi(u)\frac{1}{p^{2}-m_{c}^{2}}-
\frac{1}{16}\int_{0}^{1}du\mathcal{A}(u)(\frac{10}{(p^{2}-m_{c}^{2})^{2}}+\frac{8m_{c}^{2}}{(p^{2}-m_{c}^{2})^{3}})]\nonumber
\\
\end{eqnarray}
These results are the final results of the QCD part (OPE
expression) of the correlator. The next step is to equate Eq.
(\ref{decompositionm}) and Eq. (\ref{corr.7}) (the physical or
phenomenological side of the correlation function) and perform the
Borel transformation, with respect to the momentum of
$D_{s}^{\ast}$ meson ($p^{2}\rightarrow M_{B}^{2}$), in order to
suppress the contributions of higher states and continuum. We
obtain the following sum rules for the transition form factors,
namely:
\begin{equation}\label{sum rules}
F_{V,
A}^{(D_{s}^{\ast})}(Q^{2})=\frac{m_{D_{s}^{\ast}}}{f_{D_{s}^{\ast}}}e^{\frac{m_{D_{s}^{\ast}}}{M_{B}^{2}}}\hat{B}\{\int_{(m_{c}+m_{s})^{2}}^{s_{0}}
ds\frac{\rho_{1,2}(s, p^{2})}{s-Q^{2}}+\Pi_{1,2}^{c+d+e+f}\}
\end{equation}
where V and A are correspond to 1 and 2 in r. h. s., respectively.
In Eq. (\ref{sum rules}), in order to subtract the contributions of
the higher states and the continuum, quark-hadron duality assumption
is used, i.e. it is assumed that
\begin{equation}\label{quarkhadron}
\rho^{higher ~states}(s)=\rho^{OPE}(s)\theta(s-s_{0})
\end{equation}
In the calculations, the following rule for  the Borel
transformation is used:
\begin{equation}\label{trans}
\hat{B}\frac{1}{(p^{2}-s)^{n}}=(-1)^{n}\frac{e^{\frac{-s}{M_{B}^{2}}}}{\Gamma(n)(M_{B}^{2})^{n-1}}
\end{equation}
\section{ QCD sum rules for the form factors induced by electromagnetic penguin }
 The effective Hamiltonian for the $b\rightarrow s\gamma$ transition can be written as
  follows:
\begin{equation}\label{eff.hamil}
H=-\frac{G_{F}e}{4\pi^{2}\sqrt{2}}V_{tb}V_{ts}^{\ast}C_{7}(\mu)\overline{s}\sigma_{\mu\nu}[m_{b}\frac{1+\gamma_{5}}{2}+m_{s}\frac{1-\gamma_{5}}{2}
]bF^{\mu\nu}
\end{equation}
In order to obtain the transition amplitude, we need to calculate
the following matrix element:
\begin{equation}\label{sdmatel}
<D_{s}^{\ast}\mid\overline{s}\sigma_{\mu\nu}(1\pm\gamma_{5})q^{\nu}b\mid
B_{c}>
\end{equation}
At $q^{2}=0$, we can write this matrix element in terms of the two
gauge invariant form factors $T_{1}(0)$ and $T_{2}(0)$
\begin{eqnarray}\label{matelform}
<D_{s}^{\ast}(p,\varepsilon^{(D_{s}^{\ast})})\mid\overline{s}\sigma_{\mu\nu}q^{\nu}b\mid
B_{c}(Q)>&=&i\varepsilon_{\mu\alpha\beta\lambda}\varepsilon^{(D_{s}^{\ast})_{\alpha}}p^{\beta}Q^{\lambda}T_{1}(0),\nonumber\\
<D_{s}^{\ast}(p,\varepsilon^{(D_{s}^{\ast})})\mid\overline{s}\sigma_{\mu\nu}q^{\nu}\gamma_{5}b\mid
B_{c}(Q)>&=&[(m_{B_{c}}^{2}-m_{D_{s}^{\ast}}^{2})\varepsilon^{(D_{s}^{\ast})}_{\mu}\nonumber\\
&-&(\varepsilon^{(D_{s}^{\ast})}.q)(p+Q)_{\mu}]T_{2}(0)
\end{eqnarray}
Using the relation
\begin{equation}\label{bireind}
\sigma_{\mu\nu}\gamma_{5}=-\frac{i}{2}\varepsilon_{\mu\nu\alpha\beta}\sigma^{\alpha\beta}
\end{equation}
one can immediately obtain that $T_{2}(0)$=$\frac{1}{2}T_{1}(0)$.
Then, we need to calculate only the form factor $T_{1}(0)$. For this
aim,  we define the following three point correlation function:
\begin{equation}\label{tpcf}
\Pi_{\mu\alpha}=-\int d^{4}xd^{4}ye^{i(Qx-py)}<0\mid
T\{\overline{c}(y)\gamma_{\alpha}s(y)\overline{s}(0)\sigma_{\mu\nu}q^{\nu}b(0)\overline{c}(x)i\gamma_{5}b(x)\}
\end{equation}
where $~\overline{c}\gamma_{\alpha}s~$ and
$~\overline{c}i\gamma_{5}b~$ are the interpolating currents of
$D_{s}^{\ast}$ and $B_{c}$ mesons, respectively.

After inserting the hadrons full set with  quantum numbers of
corresponding interpolating currents (see also \cite{azizi}), we
obtain the following expression for the phenomenological part of the
correlation function:
\begin{eqnarray}\label{matelform}
\Pi_{\mu\alpha}=i\frac{f_{B_{c}}m_{B_{c}}^2}{(m_{b}+m_{c})}\frac{f_{D_{s}^{\ast}}m_{D_{s}^{\ast}}}
{(p^2-m_{D_{s}^{\ast}}^2)(Q^2-m_{B_{c}}^2)}\varepsilon_{\mu\alpha\beta\lambda}p^{\beta}Q^{\lambda}T_{1}(0)
+ \mbox{excited states.}\nonumber \\
\end{eqnarray}
For the calculation of the QCD part, we write the Lorentz
structure in the above correlator as:
\begin{eqnarray}\label{thrcor}
\Pi_{\mu\alpha}=i\varepsilon_{\mu\alpha\beta\lambda}p^{\beta}Q^{\lambda}\Pi(p^{2},Q^{2})
\end{eqnarray}
where
\begin{equation}\label{10au}
\Pi^{per}(p^{2},Q^{2})=-\frac{1}{(2\pi)^2}\int
d\tilde{s}ds'\frac{\rho^{per}(p^{2},Q^{2})}{(\tilde{s}-Q^2)(s'-p^2)}+\textrm{
subtraction terms}
\end{equation}
The standard calculations lead to the following result for the
pertubative part (bare-loop diagram):
\begin{eqnarray}\label{rhoper}
\rho^{per}(s', \tilde{s})=4N_{C}[m_{b}m_{c}(A_{1} + A_{2} + I_{0})-
m^{2}_{b} A_{1} - 2A_{3}]
\end{eqnarray}
where
\begin{eqnarray}\label{where}
A_{1} &=& \frac{2I_{0} }{(\tilde{s} - s')^{2}} [s'(\tilde{s} +
m^{2}_{c} - m^{2}_{b})
- \frac{1}{2}  (\tilde{s} + s')(s' + m^{2}_{c} )],\nonumber\\
A_{2} &=& \frac{2I_{0} }{(\tilde{s} - s')^{2}} [\frac{1}{2}
(\tilde{s} + s')( m^{2}_{b} - m^{2}_{c}-\tilde{s})
+\tilde{s}(s' + m^{2}_{c} )],\nonumber\\
A_{3}& =& I_{0} \frac{m^{2}_{b} [m^{2}_{b} s' + (m^{2}_{c} -
s')(\tilde{s} -
s')]}{ 2(\tilde{s }- s')^{2}},\nonumber\\
I_{0} &=& - \frac{1}{ 4(\tilde{s} - s')}
\end{eqnarray}
The integration regions over $\tilde{s}$ and $s'$ are obtained from
the following inequalities:
\begin{eqnarray}\label{where}
m^{2}_{c} \leq s' \leq s' _{0} ,\nonumber\\
s' - \frac{s'm^{2}_{b}}{ m^{2}_{c} - s'} \leq \tilde{s} \leq
\tilde{s_{0}}
\end{eqnarray}
The quark condensate terms give zero contribution after applying
the double Borel transformation, with respect to the $p^{2}$
($p^{2}\rightarrow M_{2}^{2}$) and $Q^{2}$ ($Q^{2}\rightarrow
M_{1}^{2}$). Only the gluon condensates can  contribute to the
form factor. Fig. 3 shows such type of diagrams.
\begin{figure}
\vspace*{-1cm}
\begin{center}
\includegraphics[width=10cm]{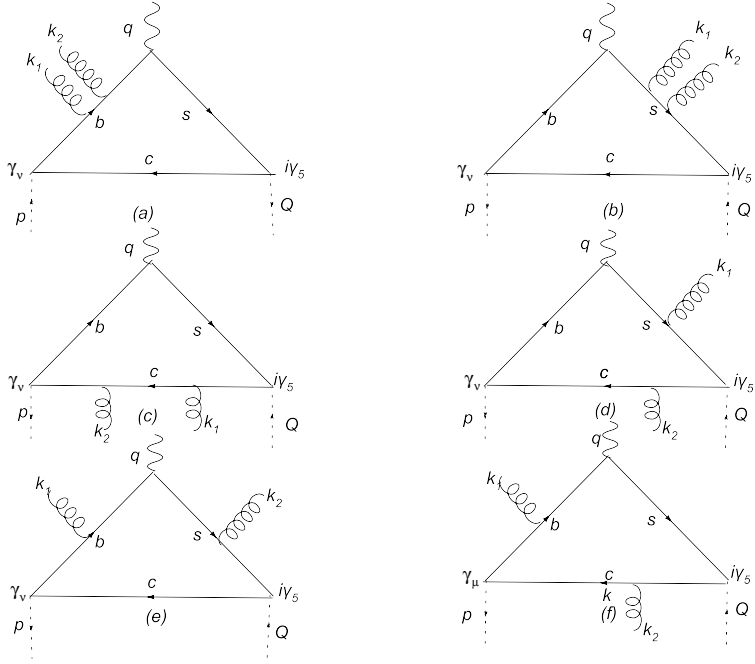}
\end{center}
\caption{Feynmen diagrams for gluon corrections } \label{fig,2}
\end{figure}
After lengthy calculations for the gluon condensates contribution
and
 equating the phenomenological and QCD parts and applying double
Borel transformation with respect to the $p^{2}$ and $Q^{2}$, we
find the following expression for the form factor $T_{1}(0)$:
\begin{eqnarray}\label{t10}
T_{1}(0)&=&-\frac{1}{(2\pi)^2}\frac{(m_{b}+m_{c})}{f_{B_{c}}m_{B_{c}}^2f_{D_{s}^{\ast}}m_{D_{s}^{\ast}}}[\int
d\tilde{s}ds'\rho^{per}(s',\tilde{s})e^{-\frac{\tilde{s}}{M_{1}^{2}}}e^{-\frac{s'}{M_{2}^{2}}}\nonumber\\
&+&M_{1}^{2}M_{2}^{2}<\frac{\alpha_{s}}{\pi}G^{2}>C_{G^{2}}]
\end{eqnarray}
where $C_{G^{2}}$ is the Wilson coefficient of the gluon
condensate and we thus have (see Fig. 3):
\begin{equation}\label{C_{G^{2}}}
C_{G^{2}}=C_{G^{2}}^{a}+C_{G^{2}}^{b}+C_{G^{2}}^{c}+C_{G^{2}}^{d}+C_{G^{2}}^{e}+C_{G^{2}}^{f}
\end{equation}
The explicit expressions for $C_{G^{2}}^{i}$ are given  below as
follows:
\begin{eqnarray}\label{C_{G^{2}}^{a}}
C_{G^{2}}^{a}&=& 96 m_b\{[m_c (I_0[1, 3, 1] + m_b^2 I_0[1, 4, 1] +
I_1[1, 3, 1] + m_b^2 I_1[1, 4, 1]  \nonumber\\
&+& I_2[1, 3, 1] - m_b^2 I_2[1, 4, 1])] + m_b (I_1[1, 3, 1] + m_b^2
I_1[1, 4, 1] \nonumber\\
&+& 2 I_3[1, 4, 1])\},
\end{eqnarray}
\begin{eqnarray}\label{C_{G^{2}}^{b}}
C_{G^{2}}^{b}&=&16 \{2 I_0[1, 1, 2] + 2 m_b m_c I_0[1, 1, 3] + 2
I_0[2, 1, 1] +
    3 m_b m_c I_0[2, 1, 2]\nonumber\\
&+&4 m_c^2 I_0[2, 1, 2] + 4 m_b m_c^3 I_0[2, 1, 3] +
    2 m_b m_c I_0[3, 1, 1] + 2 m_c^2 I_0[3, 1, 1] \nonumber\\
&+& 6 m_b m_c^3 I_0[3, 1, 2] +
    2 m_c^4 I_0[3, 1, 2] + 2 m_b m_c^5 I_0[3, 1, 3] + 2 I_1[1, 1, 2] \nonumber\\
&-& 2 m_b^2 I_1[1, 1, 3] + 2 m_b m_c I_1[1, 1, 3] + 2 I_1[2, 1, 1] -
    m_b^2 I_1[2, 1, 2] \nonumber\\
&+& m_b m_c I_1[2, 1, 2] + 4 m_c^2 I_1[2, 1, 2] -
    4 m_b^2 m_c^2 I_1[2, 1, 3] + 4 m_b m_c^3 I_1[2, 1, 3] \nonumber\\
&-&2 m_b^2 I_1[3, 1, 1] +
    2 m_c^2 I_1[3, 1, 1] - 6 m_b^2 m_c^2 I_1[3, 1, 2] + 4 m_b m_c^3 I_1[3, 1, 2] \nonumber\\
&+& 2 m_c^4 I_1[3, 1, 2] - 2 m_b^2 m_c^4 I_1[3, 1, 3] + 2 m_b m_c^5
I_1[3, 1, 3] +
    2 m_b m_c I_2[1, 1, 3] \nonumber\\
&+& I_2[2, 1, 1] + m_b m_c I_2[2, 1, 2] +
    4 m_b m_c^3 I_2[2, 1, 3] - 4 m_c^2 I_2[3, 1, 1] \nonumber\\
&+& 4 m_b m_c^3 I_2[3, 1, 2] +
    2 m_b m_c^5 I_2[3, 1, 3] - 4 I_3[1, 1, 3] - 4 I_3[2, 1, 2] \nonumber\\
&-& 8 m_c^2 I_3[2, 1, 3] - 8 I_3[3, 1, 1] - 16 m_c^2 I_3[3, 1, 2] -
    4 m_c^4 I_3[3, 1, 3]\} \nonumber\\
&-& 32 M_2^2\frac{d}{dM_2^2} \{M_2^2 [I_0[2, 1, 2] + 2 m_b m_c
I_0[2, 1, 3] +
       2 m_b m_c I_0[3, 1, 2] \nonumber\\
&+& m_c^2 I_0[3, 1, 2] + 2 m_b m_c^3 I_0[3, 1, 3] +
       I_1[2, 1, 2] - 2 m_b^2 I_1[2, 1, 3]\nonumber\\& +& 2 m_b m_c I_1[2, 1, 3]
- 2 m_b^2 I_1[3, 1, 2]+ m_b m_c I_1[3, 1, 2] + m_c^2 I_1[3, 1, 2] \nonumber\\
&-&2 m_b^2 m_c^2 I_1[3, 1, 3] + 2 m_b m_c^3 I_1[3, 1, 3]-I_2[2, 1,
2] +
       2 m_b m_c I_2[2, 1, 3] \nonumber\\
&-& 2 I_2[3, 1, 1] + m_b m_c I_2[3, 1, 2]-m_c^2 I_2[3, 1, 2] + 2 m_b
m_c^3 I_2[3, 1, 3] \nonumber\\
&-& 4 I_3[2, 1, 3] - 6 I_3[3, 1, 2] - 4 m_c^2 I_3[3, 1, 3]]\}
\nonumber\\ &-& 32 M_2^4 (\frac{d^2}{dM_2^2})^{2} \{ M_2^4 [
     m_b^2 I_1[3, 1, 3] +2 I_2[3, 1, 2]
- m_b m_c (I_0[3, 1, 3] \nonumber\\
&+& I_1[3, 1, 3] +I_2[3, 1, 3]) + 2 I_3[3, 1, 3]]
     \}~,
\end{eqnarray}
\begin{eqnarray}\label{C_{G^{2}}^{c}}
C_{G^{2}}^{c}&=&96 m_c \{-( m_b^2 m_c I_1[4, 1, 1]) +
    m_b (I_0[3, 1, 1] + m_c^2 I_0[4, 1, 1]  \nonumber\\
&+& I_1[3, 1, 1] + m_c^2 I_1[4, 1, 1] +
       I_2[3, 1, 1] + m_c^2 I_2[4, 1, 1]) - 2 m_c I_3[4, 1, 1]\},
\nonumber\\
\end{eqnarray}
\begin{eqnarray}\label{C_{G^{2}}^{d}}
C_{G^{2}}^{d}&=&-32 m_b m_c \{ I_0[2, 1, 2] + I_0[3, 1, 1] + m_c^2
I_0[3, 1, 2] + I_1[2, 1, 2] \nonumber\\
&+&  m_c^2 I_1[3, 1, 2]  + I_2[2, 1, 2] + m_c^2 I_2[3, 1, 2] - 4
I_3[3, 1, 2] \nonumber\\
&-&  M_2^2 \frac{d}{dM^2} [ M_2^2 ( I_0[3, 1, 2] + I_1[3, 1, 2] +
I_2[3, 1, 2] )] \} +16 \{m_b^2 I_1[2, 1, 2] \nonumber\\
&+&  I_2[2, 1, 1]- m_b m_c (I_0[2, 1, 2] + I_1[2, 1, 2]+ I_2[2, 1,
2]) + 4 I_3[2, 1, 2]\} ,
\nonumber\\
\end{eqnarray}
\begin{eqnarray}\label{C_{G^{2}}^{e}}
C_{G^{2}}^{e}&=&16 \{2 I_0[1, 1, 2] + I_0[1, 2, 1] + 2 m_b^2 I_0[1,
2, 2] + m_b m_c I_0[1, 2, 2]   \nonumber\\
&+&  m_c^2 I_0[2, 1, 2] + m_b m_c I_0[2, 2, 1] + m_c^2 I_0[2, 2, 1]
+
    m_b^2 m_c^2 I_0[2, 2, 2]  \nonumber\\
&+&  m_b m_c^3 I_0[2, 2, 2] + I_1[1, 1, 2] + I_1[1, 2, 1] +
    m_b^2 I_1[1, 2, 2] \nonumber\\
&+&  m_b m_c I_1[1, 2, 2] - I_1[2, 1, 1] -
    2 m_b m_c I_1[2, 1, 2] - m_b^2 I_1[2, 2, 1]  \nonumber\\
&+& m_c^2 I_1[2, 2, 1] -
    2 m_b^3 m_c I_1[2, 2, 2] + m_b m_c^3 I_1[2, 2, 2] + I_2[1, 1, 2]  \nonumber\\
&+& + I_2[1, 2, 1] +
    m_b^2 I_2[1, 2, 2] + m_b m_c I_2[1, 2, 2] - 2 m_b m_c I_2[2, 2, 1]  \nonumber\\
&+&  m_c^2 I_2[2, 2, 1] + m_b m_c^3 I_2[2, 2, 2] - 2 I_3[2, 1, 2] -
4 I_3[2, 2, 1]  \nonumber\\
&-&  2 m_b^2 I_3[2, 2, 2] - 4 m_b m_c I_3[2, 2, 2] \}  \nonumber\\
&-&  16 M_1^2\frac{d}{dM_1^2} \{ M_1^2 [
    I_2[2, 2, 1] + 2 I_3[2, 2, 2]]\}  \nonumber\\
&+&  16 M_2^2 \frac{d}{dM_2^2} \{ M_2^2[-m_b m_c I_0[2, 2, 2] +
I_1[2, 1, 2] + m_b^2 I_1[2, 2, 2]  \nonumber\\
&-&  m_b m_c I_1[2, 2, 2]  +I_2[2, 1, 2] + I_2[2, 2, 1] + m_b^2
I_2[2, 2, 2]  \nonumber\\
&-&
       m_b m_c I_2[2, 2, 2]  + 2 I_3[2, 2, 2] ] \} +  16 \{I_1[1, 1, 2] + m_b^2 I_1[1, 2, 2]  \nonumber\\
&+& I_2[1, 2, 1] - m_b m_c (I_0[1, 2, 2] + I_1[1, 2, 2]  + I_2[1, 2,
2])  \nonumber\\
&+&  2 I_3[1, 2, 2]\}~,
\end{eqnarray}
\begin{eqnarray}\label{C_{G^{2}}^{f}}
C_{G^{2}}^{f}&=&16 \{2 I_0[1, 2, 1] + 2 I_0[2, 1, 1] + 2 m_b^2
I_0[2, 2, 1]  \nonumber\\
&+&  6 m_b m_c I_0[2, 2, 1] + 2 m_c^2 I_0[2, 2, 1] + 2 I_1[1, 2, 1]
\nonumber\\
&-&
    5 I_1[2, 1, 1] - 5 m_c^2 I_1[2, 2, 1] + 6 m_b m_c I_1[2, 2, 1] +
    2 m_c^2 I_1[2, 2, 1] \nonumber\\
&+&  2 I_2[1, 2, 1] - I_2[2, 1, 1] -  m_b^2 I_2[2, 2, 1] +
    6 m_b m_c I_2[2, 2, 1] \nonumber\\
&+&  2 m_c^2 I_2[2, 2, 1] - 14 I_3[2, 2, 1] \}- 32
M_1^2\frac{d}{dM_1^2} \{ M_1^2 [I_0[2, 2, 1] \nonumber\\
&+&  I_1[2, 2, 1] + I_2[2, 2, 1] ]\}
\end{eqnarray}
and the for explicit form of the $I_{i}[a, b, c]$, we obtain:
\begin{eqnarray}\label{Io}
I_0[a,b,c] &=& \frac{(-1)^{a+b+c}}{16 \pi^2\,\Gamma(a) \Gamma(b)
\Gamma(c)} (M_1^2)^{2-a-b} (M_2^2)^{2-a-c} \,\nonumber\\&&\times
{\cal U}_0(a+b+c-4,1-c-b)~,
\nonumber\\
I_1[a,b,c] &=& \frac{(-1)^{a+b+c+1}}{16 \pi^2\,\Gamma(a) \Gamma(b)
\Gamma(c)} (M_1^2)^{2-a-b} (M_2^2)^{3-a-c} \,
\nonumber\\&&\times{\cal U}_0(a+b+c-5,1-c-b)~,
\nonumber\\
I_2[a,b,c] &= &\frac{(-1)^{a+b+c+1}}{16 \pi^2\,\Gamma(a) \Gamma(b)
\Gamma(c)} (M_1^2)^{3-a-b} (M_2^2)^{2-a-c} \,
\nonumber\\&&\times{\cal U}_0(a+b+c-5,1-c-b)~,
\nonumber\\
I_3[a,b,c] &=& \frac{(-1)^{a+b+c+1}}{32 \pi^2\,\Gamma(a) \Gamma(b)
\Gamma(c)} (M_1^2)^{3-a-b} (M_2^2)^{3-a-c} \,
\nonumber\\&&\times{\cal U}_0(a+b+c-6,2-c-b)
\end{eqnarray}
The function ${\cal U}_0(i, j)$, also, is given by:
\begin{eqnarray}\label{uo}
{\cal U}_0(i,j) = \int_0^\infty dy (y+M_1^2+M_2^2)^i y^j \,exp\left[
-\frac{B_{-1}}{y} - B_0 - B_1 y \right]~,
\end{eqnarray}
where
\begin{eqnarray}\label{b}
B_{-1} &= &\frac{m_b^2}{M_1^2} \left[M_1^2 + M_2^2 \right] ~, \cr
B_0 &=& \frac{1}{M_1^2 M_2^2} \left[M_1^2  m_c^2 + M_2^2
(m_c^2+m_b^2) \right] ~, \cr B_1 &=&\frac{m_c^2}{M_1^2 M_2^2}
\end{eqnarray}
\section{Numerical analysis}
In this section, we present our numerical analysis for the form
factors. From the sum rule expressions of these form factors, we
see that  the condensates, leptonic decay constants of $B_{c}$ and
$D_{s}^{\ast}$ mesons, continuum thresholds $s_{0}$,
$\tilde{s_{0}}$ and $s'_{0}$, the relevant parameters in photon
distribution amplitudes (DA's) and Borel parameters $M_{B}^2$,
$M_{1}^2$ and $M_{2}^2$
 are the main input parameters. In further numerical
analysis, we choose the value of the condensates at a fixed
renormalization scale of about $1$ GeV. The values of the
condensates are\cite{Ioffe}:
$<\overline{\psi}\psi\mid_{\mu=1~GeV}>=-(240\pm10~MeV)^3$,
$<\overline{s}s>=(0.8\pm0.2)<\overline{\psi}\psi>$ and
$m_{0}^2=0.8~GeV^2$.
 The quark  and mesons masses are taken to be $ m_{c}(\mu=m_{c})=
 1.275\pm
 0.015~ GeV$, $m_{s}(1~ GeV)\simeq 142 ~MeV$ \cite{Huang} ,  $m_{b} =
(4.7\pm
 0.1)~GeV$ \cite{Ioffe} ,  $m_{D_{s}^{\ast}}=2.112~GeV$
  and $ m_{B_{C}}=6.258~GeV$. For
 the values of the leptonic decay
constants of $B_{C}$ and $D_{s}^{\ast} $ mesons, we use the
results obtained from the two-point QCD analysis: $f_{B_{C}} =
0.35 GeV$ \cite{Colangelo2, Kiselev, Aliev4} and $f_{D_{s}^{\ast}}
=266\pm32
  ~MeV $\cite{Colangelo1}. The relevant parameters in
photon distribution amplitudes (DA's) are taken to be
$\chi=3.15\pm0.3
GeV^{-2},~\kappa=0.2,~\kappa^{+}=0,~\zeta_{1}=0.4,~\zeta_{1}^{+}=0,~\zeta_{2}=0.3,~\zeta_{2}^{+}=0,~f_{3\gamma}=-(4\pm2)\times10^{-3}GeV^{2}
,~w_{\gamma}^{A}=-2.1\pm1.0,~w_{\gamma}^{V}=3.8\pm1.8$ \cite{
Rohrwild, Ball3, Balitsky}. The threshold parameters are also
determined from the two-point QCD sum rules: $s_{0} =8~ GeV^2$ ,
$\tilde{s_{0}} = 45 ~GeV ^{2}$, $s'_{0} = 8 ~GeV ^{2}$
\cite{Aliev1, Colangelo1, Shifman1}. The Borel parameters
$M_{B}^2$, $M_{1}^2$ and $M_{2}^2$ are auxiliary quantities and,
therefore the results of physical quantities should not depend on
them. In the QCD sum rule method, OPE is truncated at finite
order, leaving a residual dependence on the Borel parameters. For
this reason, the working regions for the Borel parameters should
be chosen such that in these regions the form factors are
practically independent of them. The working regions for the Borel
parameters $M_{B}^2 $, $M_{1}^2 $ and $M_{2}^2 $ can be determined
on the condition that, on the one side, the continuum contribution
should be small, and on the other side, the contribution of the
operator with the highest dimension should be small. As a result
of the above-mentioned requirements, the working regions for this
transition are obtained to be:
\begin{eqnarray} &&4~ GeV^2 <
M_{B}^2 <10~ \texttt{GeV}^2, ~10\texttt{GeV}^{2}\leq
M_{1}^{2}\leq~25\texttt{GeV}^{2},\nonumber\\ &&
4\texttt{GeV}^{2}\leq M_{2}^{2}\leq 10\texttt{GeV}
^{2}.\end{eqnarray}

Now, by calculating the total decay widths  and taking $\mid
V_{cs}\mid=0.957\pm0.017$ , $\mid V_{cb}\mid=0.0416\pm0.0006$, $\mid
V_{tb}\mid=0.77^{+0.18}_{-0.24}$, $\mid
V_{ts}\mid=(40.6\pm2.7)\times10^{-3}$ \cite {Ceccucci},
$<\frac{\alpha_{s}}{\pi}G^{2}>=0.012 GeV ^{4}$ \cite{Shifman1},
$C_{7}(\mu=m_{c})= -0.0068 - 0.02i$ \cite {Alievsp} and
$\tau_{B_{c}}=0.52 \times10^{-12}s$ \cite{Beneke}, we obtain the
numerical results of the electromagnetic penguin(EP), weak
annihilation(WA) and total branching ratios for this decay as
follows:

\begin{eqnarray}\label{branching1}
&&\textbf{B}^{(EP)}(B_{c}\rightarrow
D_{s}^{\ast}\gamma)=3.468\times10^{-6}\nonumber\\
&&\textbf{B}^{(WA)}(B_{c}\rightarrow
D_{s}^{\ast}\gamma)=1.557\times10^{-5}\nonumber\\
&&\textbf{B}^{(Total)}(B_{c}\rightarrow
D_{s}^{\ast}\gamma)=2.462\times10^{-5}
\end{eqnarray}
 From the above results, we see that the weak annihilation contribution to the
 total branching ratio is about $4.48$ times greater than that
 of the electromagnetic penguin diagram. Here, it is observed that the difference
 between  the total branching ratio with sum  of  the weak
  annihilation and electromagnetic penguin branching ratios
   comes from the cross term in total decay width.  Also our result for the total
 branching ratio shows that the $B_{c}\rightarrow D_{s}^{\ast}\gamma$
  decay can be  measured  at LHC.

Now, we compare our results of the $B_{c}\rightarrow
D_{s}^{\ast}\gamma$ to the results of the perturbative QCD
\cite{Dongsheng}, relativistic independent quark model \cite{Barik},
pertubative QCD in standard model (SM (PQCD)) \cite{Gongru} , multi
scale walking technicolor (MWTCM) \cite{Gongru}  and topcolor
assisted MWTCM (TAMWTCM) \cite{Gongru} for $\tau_{B_{c}}=0.52
\times10^{-12}s$ö as shown in Table (1).
\begin{table}[h]
\centering
\begin{tabular}{|c|c|c|c|} \hline
  & $\textbf{B}^{EP}(B_{c}\rightarrow D_{s}^{\ast}\gamma)$  & $\textbf{B}^{WA}(B_{c}\rightarrow D_{s}^{\ast}\gamma)$
   & $\textbf{B}^{Total}(B_{c}\rightarrow D_{s}^{\ast}\gamma)$\\\cline{1-4}
 Present study & $3.468\times10^{-6}$& $1.557\times10^{-5}$& $2.462\times10^{-5}$\\\cline{1-4}
 PQCD & $3.70\times10^{-6}$  & $4.94\times10^{-6}$ & $1.14\times10^{-5}$\\\cline{1-4}
 RIQM &$ 2.40\times10^{-5} $&$4.51\times10^{-5}$& $1.39\times10^{-4}$\\\cline{1-4}
 MWTCM &$(0.68-3.42)10^{-4}$& $(0.74-0.81)10^{-6}$ & $(0.74-3.57)\times10^{-4}$\\\cline{1-4}
  TAMWTCM & $(5.18-7.23)10^{-7}$& $(7.24-8.13)10^{-7}$ &$(1.78-9.95)\times10^{-6}$ \\\cline{1-4}
   SM(PQCD) &$1.73\times10^{-7} $& $5.89\times10^{-7}$& $7.83\times10^{-7}$\\\cline{1-4}
 \end{tabular}
 \vspace{0.8cm}
\caption{Comparison of the branching ratio for  $B_{c}\rightarrow
D_{s}^{\ast}\gamma$ decay based on the $\tau_{B_{c}}=0.52
\times10^{-12}s$.} \label{tab:1}
\end{table}

 Looking at this table, it is  seen that there is a good
agreement  between the present study and the PQCD \cite{Dongsheng},
 in order of magnitude for the total branching ratio. However, our
result is approximately one order of magnitude less than that of the
RIQM and MWTCM. Also, it is one order of magnitude and two orders of
magnitude greater than that of the TAMWTCM and  SM(PQCD)
\cite{Gongru}, respectively. The ratio of
$\textbf{B}^{WA}/\textbf{B}^{EP}$ for the present work, PQCD
 \cite{Dongsheng},  RIQM,
  SM(PQCD) \cite{Gongru}, TAMWTCM, MWTCM are 4.48, 1.34, 1.9, 3.4,
  1.23 and 0.01,
respectively. As a result of the above discussions, we can say that
in the QCD sum rules (present study), relativistic independent quark
model, perturbative QCD and TAMWTCM approaches, the weak
annihilation contribution to the total branching ratio dominates the
contribution coming from the electromagnetic penguin diagram, but
this is not true  only for the MWTCM approach. The presence of the
pseudo Goldstone bosons in the MWTCM leads to  a  discrepancy
between this model and the other two models in \cite{Gongru} (for
more details see\cite{Gongru}) and a part of inconsistency in the
results of the different methods may be related to the different
magnitudes of the input parameters, getting from different
references; e.g. we use $ m_{c}(\mu=m_{c})=
 1.275\pm
 0.015~ GeV$ for the c quark masses while the authors of \cite{Gongru} use $ m_{c}=1.6~GeV$ and also to the nature
 of the methods and their accuracy.

In this step, for the analysis of $B_{c}\rightarrow D^{\ast}\gamma$,
in the entire calculations we replace the s quark with the d quark.
Making $m_{D_{s}^{\ast}}\rightarrow m_{D^{\ast}}$,
$f_{D_{s}^{\ast}}\rightarrow f_{D^{\ast}}$, $V_{ts}\rightarrow
V_{td}$, $V_{cs}\rightarrow V_{cd}$ changes and taking
$m_{D^{\ast}}=2.010~GeV$, $f_{D^{\ast}}=0.23\pm0.02~GeV$
\cite{Bowler}, $V_{cd}=0.230\pm0.011$,
$V_{td}=(7.4\pm0.8)\times10^{-3}$ \cite {Ceccucci} and $m_{d}=5~MeV$
we obtain the numerical results as below:
\begin{eqnarray}\label{branching2}
&&\textbf{B}^{(EP)}(B_{c}\rightarrow
D^{\ast}\gamma)=1.151\times10^{-7}\nonumber\\
&&\textbf{B}^{(WA)}(B_{c}\rightarrow
D^{\ast}\gamma)=2.162\times10^{-6}\nonumber\\
&&\textbf{B}^{(Total)}(B_{c}\rightarrow
D^{\ast}\gamma)=2.786\times10^{-6}
\end{eqnarray}
These results also enhance the importance of the weak annihilation
contribution to the total branching ratio in comparing with the
electromagnetic penguin diagram ones for the $B_{c}\rightarrow
D^{\ast}\gamma$. Finally, we compare our results to the relativistic
independent quark model (RIQM) \cite{Barik} for $\tau_{B_{c}}=0.52
\times10^{-12}s$ in Table
(2).\\
\begin{table}[h]
\centering
\begin{tabular}{|c|c|c|c|} \hline
  & $\textbf{B}^{EP}(B_{c}\rightarrow D^{\ast}\gamma)$  & $\textbf{B}^{WA}(B_{c}\rightarrow D^{\ast}\gamma)$
   & $\textbf{B}^{Total}(B_{c}\rightarrow D^{\ast}\gamma)$\\\cline{1-4}
 Present study & $1.151\times10^{-7}$& $2.162\times10^{-6}$& $2.786\times10^{-6}$\\\cline{1-4}
RIQM \cite{Barik}&$5.70\times10^{-7}$  &$1.33\times10^{-6}$&
$3.64\times10^{-6}$\\\cline{1-4}
 \end{tabular}
 \vspace{0.8cm}
\caption{Comparison of the branching ratio for  $B_{c}\rightarrow
D^{\ast}\gamma$ decay based on the $\tau_{B_{c}}=0.52
\times10^{-12}s$.} \label{tab:1}
\end{table}
From the Table 2, it is also seen a  good agreement  in the order of
magnitude between the present study and the relativistic independent
quark model.

 In conclusion, the present study concentrated on the radiative $B_{c}\rightarrow D_{s}^{\ast}\gamma $ and $B_{c}\rightarrow D^{\ast}\gamma $
 decays  in the framework of QCD sum rules.
  The form factors responsible
for these decays were calculated.
  The branching ratio for this decays  were estimated. The results show that
  the $B_{c}\rightarrow D_{s}^{\ast}\gamma $ case can  be measured at  LHC in
the near future.
\section{Acknowledgment}
One of the  authors (K. Azizi) would like to thank TUBITAK, Turkish
Scientific and Research Council, for their partially support. Also,
V. Bashiry would like to thank theory group of CERN for their
hospitality.

\clearpage
\end{document}